\documentclass[useAMS,usenatbib,usegraphicx]{mn2e}
\newcommand{\uv}{\mbox{$u$-$v$}}

\newcommand{\pyear}{\mbox{\% yr$^{-1}$}}
\newcommand{\kms}{\mbox{km s$^{-1}$}}
\newcommand{\Jb}{\mbox{Jy bm$^{-1}$}}
\newcommand{\GX}{\mbox{G21.5$-$0.9}}
\newcommand{\PSRJ}{\mbox{PSR~J1833$-$1034}}
\newcommand{\Wmsr}[1]{\mbox{$\, \times ~10^{#1}$} {\mbox W~m$^{-2}$ Hz$^{-1}$ sr$^{-1}$}}

\newcommand{\tablenotemark}[1]{$^{\rm #1}$}
\newcommand{\tablenotetext}[2]{\noindent$^{\rm #1}$ #2}

\title[Expansion and Radio Spectral Index of G21.5--0.9]{The Expansion
and Radio Spectral Index of G21.5--0.9: Is \PSRJ\ the Youngest Pulsar?}

\author[Bietenholz \& Bartel]{M. F. Bietenholz$^{1,2}$ and N. Bartel$^2$\\
$^1$Hartebeesthoek Radio Observatory, PO Box 443, Krugersdorp,
1740, South Africa \\
$^2$Department of Physics and Astronomy, York University, Toronto,
M3J~1P3, Ontario, Canada \\}

\begin{document}

\date{\today; Accepted for publication in the Mon.\ Not.\ R.\ Astron.\ Soc.}
\pagerange{\pageref{firstpage}--\pageref{lastpage}} \pubyear{2007}


\maketitle

\label{firstpage}

\begin{abstract}
We report on new 5-GHz VLA radio observations of the pulsar-powered
supernova remnant \GX.  These observations have allowed us to make a
high-quality radio image of this remnant with a resolution of
$\sim$0.7\arcsec.  It has a filamentary structure similar to that seen
in the Crab Nebula.  Radio structure suggestive of the torus seen
around the Crab pulsar is tentatively identified. We also compared the
new image with one taken $\sim$15~yr earlier at 1.5~GHz, both to find
the expansion speed of the remnant and to make a spectral index image.
Between 1991 and 2006, we find that the average expansion rate of the
remnant is $0.11 \pm 0.02$\pyear, corresponding, for a distance of
5~kpc, to a speed of $910 \pm 160$~\kms\ wrt.\ the centre of the
nebula.
Assuming undecelerated expansion, this expansion speed implies that
the age of \GX\ is $870^{+200}_{-150}$~yr, which makes \PSRJ\ one of
the youngest, if not the youngest, known pulsars in the Galaxy.
\end{abstract}


\section{Introduction}

The supernova remnant \GX\ (SNR 021.5$-$00.9)
has been known for over 30 years \citep[e.g.,][]{Altenhoff+1970}, and
has long been classified as a filled-centre or a Crab-like remnant
\citep[e.g.,][]{WilsonW1976}.
Such supernova remnants are powered by a central pulsar rather than by
the interaction of the expanding ejecta shell with its surroundings.
We will refer to them using the term ``pulsar wind nebula'' or PWN.~
\GX\ is a bright, centrally condensed radio and X-ray source, with a
diameter of $\sim$1\arcmin.  A larger, low surface brightness halo, of
diameter $\sim$2.5\arcmin, is seen in the X-ray \citep{Bocchino+2005,
MathesonS2005, Safi-Harb+2001, Slane+2000}.  The distance to \GX\ is
$\sim$5~kpc \citep{Camilo+2006, DavelaarSB1986}.  Its radio and X-ray
luminosities are $\sim$10\% and $\sim$1\% of those, respectively, of the
Crab Nebula.

However, despite numerous searches, no pulsar was seen in \GX\ until
very recently, when two teams announced the discovery of \PSRJ\
\citep{Gupta+2005, Camilo+2006}.  The pulsar is very faint,
with a pulse-averaged flux density of $\sim 70\, \mu$Jy at 1.4~GHz.
It has a period of 61.8~ms and a period derivative of $\dot P =
2.02\times10^{-13}$, giving it a characteristic age of 4800~yr.
Although no other relatively direct measurements of \GX's age have
been made, various arguments have led other authors to suggest an age
of $\sim$1000~yr \citep{BockWD2001,
Bocchino+2005,Camilo+2006}. Despite its faint pulsed emission, the
pulsar is in fact very energetic, having a spindown luminosity of
$\dot E = 3.3\times10^{37}$~erg~s$^{-1}$, which, in our Galaxy, is
second only to that of the Crab pulsar.

The powerful winds of young pulsars with high spindown luminosities
have been shown to excite complex structures, visible from the X-ray
to the radio \citep[e.g.,][]{GaenslerS2006}.  In particular, in the
Crab Nebula, narrow emission features, called ``wisps'' are seen.  The
wisps are thought to be associated with the shock in the pulsar wind,
and with a torus seen in the X-ray.  They move rapidly outward and are
seen in the radio, optical and X-ray \citep[e.g.,][and references
therein]{Crab-2001, Crab-2004}. 3C~58, which is
also the wind nebula of a young, energetic pulsar, exhibits a narrow radio
feature which is probably associated with the pulsar outflow,
although it is not as mobile as the Crab wisps \citep{FrailM1993,
3C58-2006}.

While numerous radio images of \GX, at frequencies from 330~MHz to
94~GHz, have been published \citep{BeckerK1976, WilsonW1976,
BeckerS1981, MorsiR1987, Furst+1988, Kassim1992, BockWD2001}, none of
them had a resolution higher than 4\arcsec, which is not sufficient to
see such wisps and other radio features in detail.
\citet{FrailM1993} observed \GX\ with the VLA specifically to search
for radio emission associated with the (then still unknown) pulsar,
but found none.  Now the pulsar location is known, it seemed well
worthwhile to revisit \GX\ with deeper and higher-resolution radio
observations.  We therefore observed \GX\ using the NRAO\footnote{The
National Radio Astronomy Observatory, NRAO, is a facility of the
National Science Foundation operated under cooperative agreement by
Associated Universities, Inc.}
Very Large Array (VLA) with the goals of making a sub-arcsecond
resolution radio image, searching for features near the pulsar, and,
by comparing our observations to older ones, determining the expansion
rate, age, and spectral index distribution of the synchrotron nebula.
We describe the observations in \S~\ref{sobs}, present the new image
in \S~\ref{simg}, determine the expansion rate and form a spectral
index image in \S~\ref{sexpand}, and discuss our findings in
\S~\ref{sconclude}.

\section{Observations and Data Reduction}
\label{sobs}

We observed \GX\ at 5~GHz, using the A and B array configurations of
the VLA. All our observations were phase-referenced to the radio
source PMN J1832$-$1035.  The details of
the observations are given in Table~\ref{tobs}, which also describes
the two archival data sets that we use in this paper.  The flux
density calibration for all observing sessions was done by observing
3C~286.  The data reduction was carried out using standard procedures
from NRAO's AIPS software package, which were also used to re-reduce
the archival data.

\begin{table*}
\caption{Observing Sessions for \GX}
\begin{tabular}{l c c c}
\hline
\multicolumn{1}{c}{Date} 
               & \multicolumn{1}{c}{Frequency (GHz)}
               & \multicolumn{1}{c}{Array Configuration}
               & \multicolumn{1}{c}{Observing Time\tablenotemark{b} (hrs)} \\
\hline
2006 Mar 7  & 4.93, 4.56 & A & 6 \\
2006 Sep 9  & 4.93, 4.56 & B & 2 \\
1991 Jul 3\tablenotemark{c}\tablenotemark{d}  & 1.46, 1.51 & B & 6 \\
1985 Nov 30\tablenotemark{c} & 4.89, 4.84 & D & 1.5 \\
\hline
\end{tabular}
\medskip
\parbox{4.5in}{
\tablenotetext{a}{The sky frequencies of the two intermediate
frequencies (IF) used.  In each case the bandwidth per IF was 50 MHz.} \\
\tablenotetext{b}{The approximate total observing time, including
calibrator observations; note that for 1985 Nov 30, only $\sim$10~min
were spent on-source for \GX.} \\
\tablenotetext{c}{Data from the NRAO data archive} \\
\tablenotetext{d}{Original results published in \citet{FrailM1993}.}}
\label{tobs}
\end{table*}

To make the best possible image, we used maximum entropy
deconvolution, which is better suited to extended objects like \GX\
than the more common CLEAN deconvolution.  However, since the diameter
of the PWN is $\sim$1\arcmin, even the shortest spacings with the
B-array recover only around half the total flux density.  Therefore
the large-scale structure may not be well recovered without some
additional short-spacing information.  We used an archival 5~GHz VLA
data set taken in the D configuration on 30 Nov 1985 to supply
additional short spacing information.

\section{High Resolution Image}
\label{simg}

\begin{figure*}
\centering \includegraphics[width=\textwidth, trim=0 0.0in 0 0.0in,
clip]{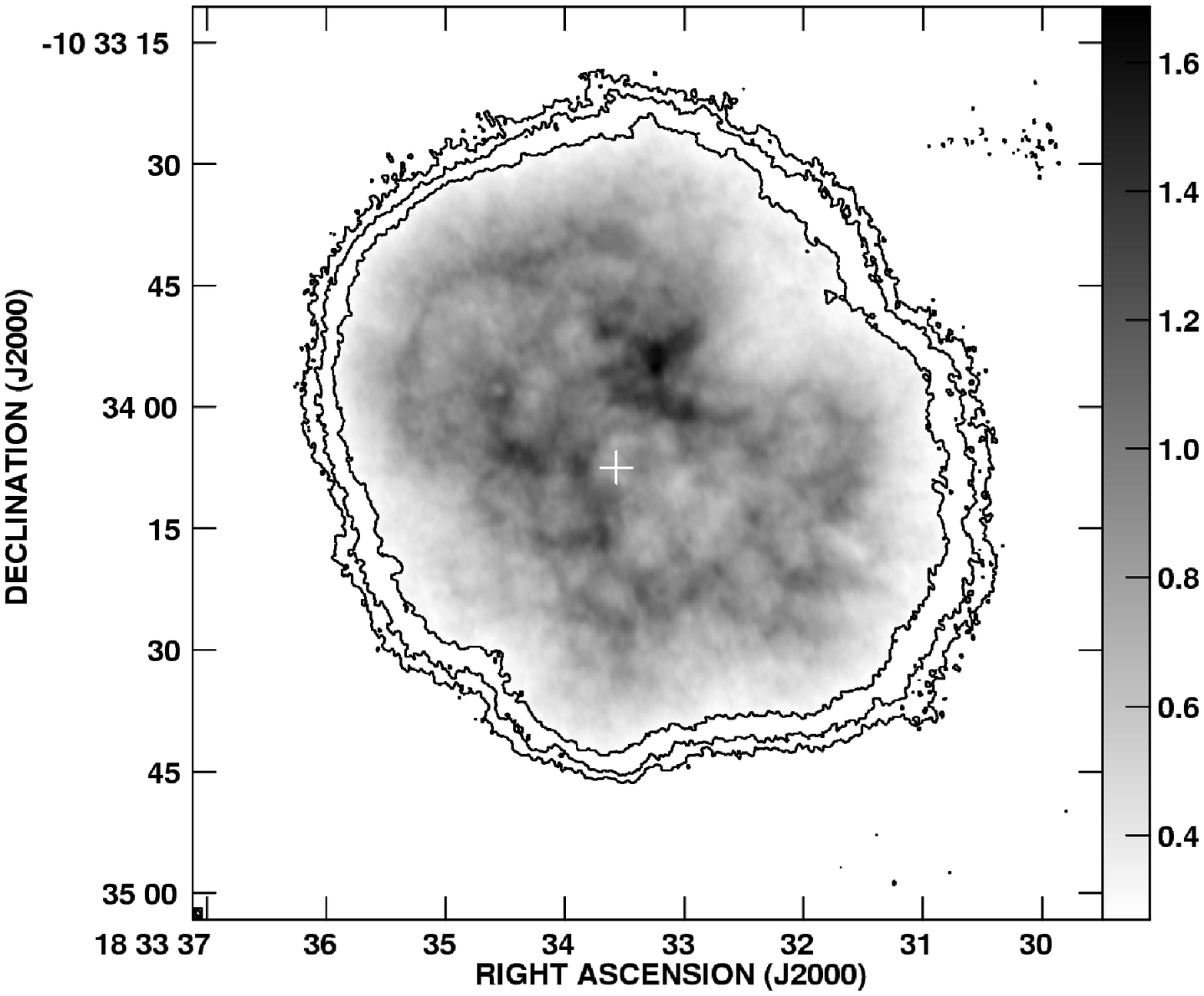}
\caption{A radio image of \GX\ at 4.75~GHz taken on 2006 March 3.  The
FWHM of the elliptical Gaussian restoring beam was $0.82 \times
0.53$\arcsec\ at p.a.\ 10\degr.  Contours are drawn at $-4$, 4, 8 and
16\% of the peak brightness of 1.69~m\Jb\ and the rms background
brightness was 19~$\mu\Jb$.  The grey scale is labelled in m\Jb.  The
white cross marks the position of the pulsar, taken from
\citet{Camilo+2006}, and known to an accuracy of $<0.5\arcsec$.  The
image was made from A and B array configuration data taken in 2006,
with the addition of a small amount of archival D array configuration
data from 1985 (see text, \S~\ref{sobs}).  In the electronic edition,
the accompanying animation shows alternating images from 2006 and
1991, visually showing the expansion over the 15-year period.  The two
images are on the same scale, and the larger and smoother of the two
is that from 2006, with the smoother appearance being due to the
higher signal-to-noise (animated gif file available from {\tt
http://lanl.arxiv.org} by downloading ``source'' from ``other
formats'').
\label{fg21image}
}
\end{figure*}

In Figure~\ref{fg21image} we show the full radio-image of G21.5,
obtained by combining the A, B, and D array observations at
5~GHz\footnote{In anticipation of our result on the expansion,
described in \S~\ref{sexp} below, the D-configuration data, taken 21
years before the A and B-configuration data, was scaled up by 2.3\%,
based on our measurement of the expansion speed.  The effect on the
image of this small correction to the D-configuration data is
minimal.},
and using a AIPS' implementation of the Briggs robustness parameter
\citep{BriggsSS1999} of 1.5, which achieves a resolution somewhat
better than available through natural weighting with only slightly
increased noise.  The full-width at half-maximum (FWHM) of the
elliptical Gaussian restoring beam was $0.82\arcsec \times
0.53\arcsec$ at p.a.\ 10\degr.  The total flux density recovered was
$6.7 \pm 0.3$~Jy, consistent
with the earlier total flux density measurements at cm wavelengths
\citep{WilsonW1976, Salter+1989}.  The off-source rms background
brightness was 19~$\mu\Jb$.  This image has notably higher resolution
and lower rms background than any previously published radio images of
\GX.  Prominent filamentary structure, reminiscent of that seen in
radio images of the Crab nebula \citep{Crab-2001, Crab-2004} and 3C~58
\citep{3C58-2006, 3C58-2001}, is clearly visible.

The remnant has a double-lobed structure, with the axis of symmetry
running approximately from the northwest to the southeast.  This
structure has also been noted in the X-ray
\citep[e.g.,][]{MathesonS2005,Safi-Harb+2001}.  Unlike in the X-ray,
however, there is no bright central condensation in the radio, and the
northwest lobe is somewhat brighter than the southeast one.

The position of the pulsar is known to within $<0.5\arcsec$ from
{\em Chandra} X-ray images \citep{Camilo+2006}\footnote{We note
that the position of the pulsar itself has not been accurately measured,
but that the position of the bright, although un-pulsed, central condensation
in X-rays is identified as the pulsar position by \citet{Camilo+2006}.}.
Although the pulsar is too faint to discern in our radio images, we
can accurately locate it, since our observations were phase-referenced
to the radio calibrator PMN J1832$-$1035.
We estimate the uncertainty of our astrometric calibration is also
about 0.5\arcsec.  The pulsar position is indicated in
Fig.~\ref{fg21image}.  The pulsar is displaced by $\sim$4\arcsec\
southeast of the centroid of the remnant.  Loop-like filaments are
seen, with the most prominent being to the northwest.

\begin{figure}
\centering \includegraphics[width=0.5\textwidth, trim=0 0.0in 0 0.0in,
clip]{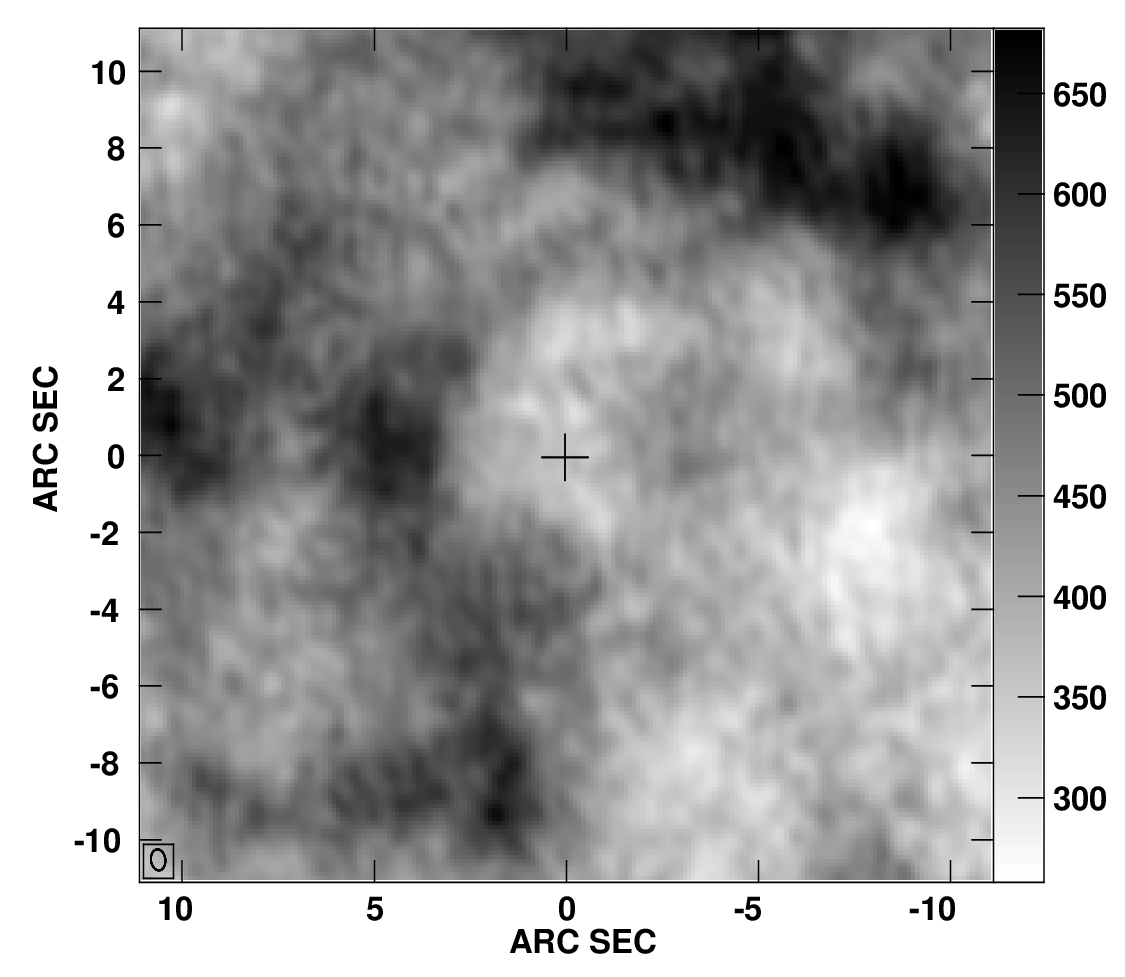}
\caption{A detail image of the region near the pulsar, made from the
same data but at a slightly higher resolution than that of
Figure~\ref{fg21image}.  The axes show the RA and decl.\ offsets from
the pulsar position, which is marked with a cross.  The FWHM of the
elliptical Gaussian restoring beam was $0.55 \times 0.37$\arcsec\ at
p.a.\ 7\degr.  The grey scale is labelled in $\mu\Jb$.
\label{fcenter}
}
\end{figure}

We show a detail of the region near the pulsar, made with a slightly
higher resolution (AIPS robustness parameter $= -1$) of $0.55\arcsec
\times 0.37\arcsec$ at p.a.\ 7\degr\ in Fig.~\ref{fcenter}.  The
brightest region is $\sim$10\arcsec\ north-northwest of the
pulsar. The pulsar itself is in a region of somewhat lower surface
brightness of diameter $\sim$8\arcsec, which is open to the southeast.
There is possibly an elliptical structure, incomplete to the
southeast, oriented at p.a.\ 45\degr\ and of major axis diameter
$\sim$8\arcsec\ (corresponding to $6\times10^{17}$~cm at 5~kpc),
although a conclusive identification is not possible at this
combination of signal-to-noise and resolution.  If the structure is
real, then its minor axis is oriented along the axis of symmetry of
the nebula (i.e., northwest to southeast).

\section{Expansion Speed and Spectral Index}
\label{sexpand}

We set out to both determine \GX's rate of expansion and to map its
radio spectral index by comparing our 5~GHz image from 2006 to a
1.5~GHz one made from data taken in 1991 in the A array configuration
\citep[see][for the original results from the 1991 data]{FrailM1993}.
Although an archival image at our observing frequency would have been
preferable for determining the expansion, no older 5-GHz image of
sufficient sensitivity and resolution was available.  Since the two
images differ in both epoch and frequency, the expansion and spectral
index cannot be determined independently of each other.  However, for
a source such with well-resolved, small-scale structure such as \GX,
and for a range in frequency of only three, it is highly unlikely that
variation in the spectral index would cause the source to appear
larger or smaller at different frequencies. Furthermore, earlier
results on \GX\ \citep{MorsiR1987, Furst+1988} suggest that \GX's
radio spectral index is relatively uniform across the nebula.  Large
variations in spectral index are not in fact expected in PWNe, if the
remarkably uniformity of spectral index over each of the two most
prominent remnants, the Crab Nebula and 3C~58, is any guide
\citep{Crab-1997, 3C58-2001}.  Therefore, since the expected
variations in spectral index are small, and very unlikely to mimic a
frequency-dependent change size, we proceed to compare the 2001
1.5-GHz and 2006 5-GHz images to determine the expansion rate.  Once
the expansion rate is known, we can ``expand'' the 1991 1.5-GHz image
in order to make a spectral-index image.

\subsection{Expansion}
\label{sexp}

We used the same approach to determining the expansion as was used in
\citet{3C58-2006} and \citet{3C58-2001, Crab-expand}, and repeat a
brief description here for the convenience of the reader.  The goal is
to determine the overall or average expansion rate of the radio
nebula.  Since there are few well-defined compact features, the
expansion is measured not by determining the proper motion of
individual features, but by determining an overall scaling between a
pair of images by least-squares. This was accomplished by using the
MIRIAD \citep{SaultTW1995} task IMDIFF\footnote{As a check we also
re-implemented the IMDIFF algorithm as an AIPS script.  The results
from this re-implementation, which uses different image-interpolation
and minimisation and schemes, were consistent to within our
uncertainties.}
which determines how to make one image most closely resemble another,
by calculating unbiased estimators for the scaling in size, $e$, the
scaling and the offset in flux density, $A$ and $b$ respectively, and
the offsets in RA and decl., $x$ and $y$ respectively, by least
squares.  Our chief interest is in the expansion factor, $e$, but
because of uncertainties in flux calibration, absolute position, and
image zero-point offsets caused by missing short spacings, all five
parameters needed to be determined.  This method was originally
developed by \citet{TanG1985}.

In order not to contaminate our results on the expansion, we used an
image made from the 2006 data set without the added D-configuration
data from 1985.  Although the structure at relatively small spatial
scales, namely the filamentary structure visible in
Fig.~\ref{fg21image}, was well sampled by our A and B-configuration
VLA observations, the largest scale structure was less well sampled,
and may thus not have been sufficiently well recovered by the
deconvolution.  We will therefore obtain more reliable results on the
expansion if we use high-pass filtered images, which exclude the
poorly sampled structure on the largest scales. To make the images for
determining the expansion, we performed an initial high-pass filtering
by using only visibility data at \uv~distances $>6$~K$\lambda$.  Then,
we used CLEAN deconvolution, since the positivity constraint applied
in MEM deconvolution is not applicable to such high-pass filtered
images.  Finally, the resulting CLEAN images were further high-pass
filtered in the image plane by applying a Gaussian high-pass filter of
FWHM 15\arcsec, and then the region exterior to \GX\ was
blanked\footnote{We use a slightly modified version of IMDIFF which
treats blanked regions correctly}.
The 1991 1.5-GHz data were re-reduced in a manner consistent with the
2006 data, and a similarly high-pass filtered image was made.  The FWHM
resolution of this data set was $1.5\arcsec \times 1.2\arcsec$ at
$-5$\degr, the peak brightness was 6.2~m\Jb, and the rms background 
was 0.12~m\Jb.  Before running IMDIFF, the 2006 image was convolved to
the resolution of the 1991 one.  The fitting region encompassed
$\sim$2.8 square arcminutes or $\sim$5000 beam areas.

The fitted value of the expansion parameter, $e$, between epoch 1991.5
and 2006.3 was 1.017, in other words, the nebula has expanded by 1.7\%
over the 14.8~yr period between the observations.  The rms residual
to the expansion fit was $\sim$0.11~m\Jb, which is comparable to the
combined background rms in the two images.

The uncertainty of the expansion parameter is difficult to estimate
reliably.  The purely statistical uncertainty will be small due to the
large number of image elements ($\sim 5000$ beam areas).  However, the
image elements are not strictly independent, so the expansion
uncertainty is likely dominated by systematics such as deconvolution
errors.  We estimate our uncertainty from the largest differences in
the value of the expansion parameter, $e$, observed between IMDIFF and
the AIPS minimisation and between different levels of high-pass
filtering.  Our uncertainty is therefore not a statistical $1\sigma$
value, but rather an estimated $1\sigma$ confidence interval.

Since the two images are taken at different frequencies, caution is
warranted when comparing them.  However, as mentioned, we think it
highly unlikely that any possible spectral index variations might
cause a spurious apparent change in size with frequency for such a
complex source structure.  We therefore believe our measured expansion
factor accurately reflects true spatial expansion of the nebula.
Expansion
by a factor of $1.7 \pm 0.3$\% over a 14.8~yr period suggests an
expansion rate of $0.11 \pm 0.02$~\pyear, and in the absence of any
acceleration or deceleration, an age of $870^{+200}_{-150}$~yr.
We note here that a small amount of acceleration, and therefore
a slightly older age is expected.  We discuss this subject further
in \S~\ref{sconclude} below.

\subsection{Spectral Index Image}
\label{sspix}

\begin{figure*}
\centering \includegraphics[width=\textwidth, trim=0 0.0in 0 0.0in, clip]{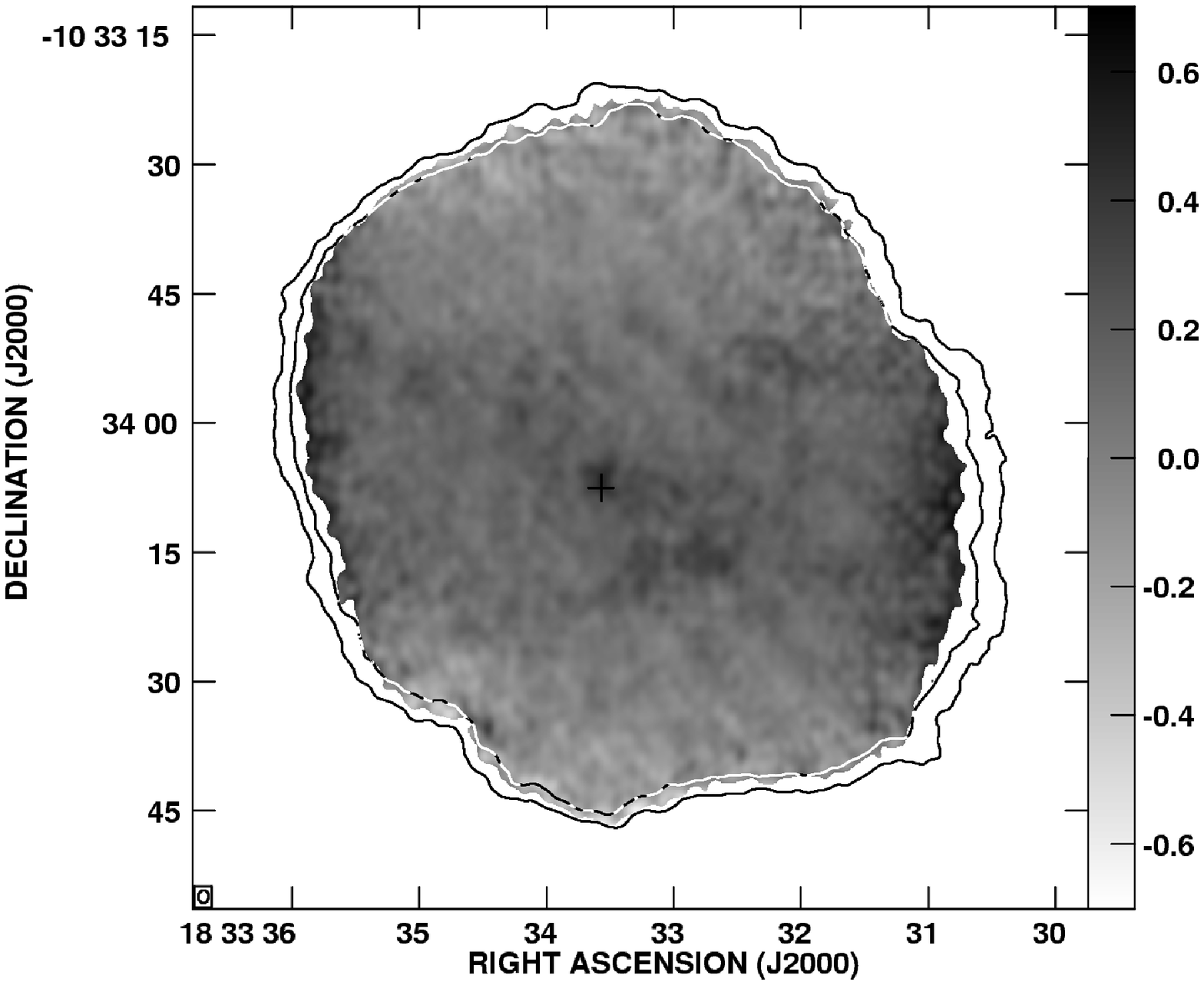}
\caption{An image of the radio spectral index of \GX\ between 1.5 and
5~GHz. The spectral index ($S_\nu \propto \nu^\alpha$) is plotted in
greyscale, and regions where the uncertainty in the spectral index
were larger than 0.2 are blanked.  The FWHM resolution is $1.5 \times
1.2$\arcsec\ at p.a.\ $-5$\degr. The two contours show the 5 and
10\% contours of the image at 5~GHz for reference (note that this
image is at lower resolution than the one in Fig.~\ref{fg21image}).
The cross again denotes the pulsar position.
\label{fspix}
}
\end{figure*}

Having established the expansion rate, we can correct the 1.5~GHz
image for the expansion between 1991 and 2006, and then form a
spectral index image with the two images, i.e., between 1.5 and
4.9~GHz.  For the spectral index image, we used different images than
for the expansion which are filtered neither in the visibility or the
image-plane.  The 5-GHz image was made incorporating the small amount
of D-configuration data from 1985, after scaling of the latter, as
described in \S~\ref{sobs} above in order to recover the large-scale
structure as accurately as possible.  We show our spectral index image
in Fig.~\ref{fspix}.  The average spectral index over our image (with
spectral index, $\alpha$, defined so that $S_\nu \propto \nu^\alpha$)
was\footnote{The uncertainty is the estimated
systematic uncertainty due to flux calibration and a possibly slightly
incomplete recovery of the total flux density in the 1.5~GHz image}
$+0.08^{+0.06}_{-0.09}$.
The rms of $\alpha$ over the image was 0.14.  The average value of
$\alpha$ is consistent with the earlier integrated values of $0$
\citep{MorsiR1987} and $-0.02$ \citep{Salter+1989}.

The spectral index is relatively uniform over the nebula. Although a
fairly large range of spectral index is visible in the image, a large
uncertainty in the spectral index is unavoidable in regions of
relatively low signal-to-noise. The formal uncertainty in the spectral
index is, on average, 0.1 over the central $\sim$40\arcsec\ of the
nebula, and as there are many image elements, a number of deviations
exceeding $2\sigma$, i.e., 0.2 in $\alpha$, are to be expected merely
because of the noise.  The regions showing larger deviations from the
mean spectral index are near the edge of the nebula where the
uncertainty is higher.  In addition, a wide band of slightly steeper
spectral index ($\Delta\alpha \sim 0.15$) running almost east-west is
seen across the center of the nebula.  This ``feature'' is almost
certainly an artefact due to slight deconvolution errors due to
incomplete sampling near the center of the \uv~plane.

No systematic deviations from the mean spectral index are visible near
the edge of the nebula, with a the mean spectral index of a strip of
width $\sim$10\arcsec\ around the circumference of the nebula differing
only insignificantly (by $-0.03$) from the overall mean.
No strong deviations are seen near the filamentary structures.  The
region near the pulsar has an $\alpha$ larger (i.e., the spectrum is
more inverted) than that of its surroundings by $\sim+0.1$ and the
brightest region has a slightly smaller $\alpha$ (i.e., the spectrum
is slightly steeper).  Although both of these deviations may be real,
they are not larger than the uncertainties due to the noise and
deconvolution errors.

\section{Discussion}
\label{sconclude}

We have made new, high-resolution 5-GHz radio images of the pulsar
wind nebula, \GX, whose pulsar was only recently detected and found to
have a very high spindown luminosity.  By comparing our image to an
earlier image at 1.5~GHz we determined both the expansion rate and the
radio spectral index of the nebula .

Our image clearly shows that \GX\ has a filamentary structure in the
radio, similar to that seen in radio images of the Crab and of 3C~58.
Large loop-like structures are seen especially towards the northeast
and southwest.  Such structure seems to be present in all young PWNe.
\citet{Slane+2004} suggested that they were loops torn by the kink
instability from the toroidal field produced by the pulsar.

The nebula is expanding fairly rapidly: our estimate of the expansion
rate between 1991 and 2006 is $0.11 \pm 0.02$\pyear.  At 5~kpc this
rate corresponds to a speed (wrt.\ the centre) of $910 \pm
160$~\kms. This is similar to although a little slower than the
expansion speeds seen for two other young PWNe: the Crab Nebula and
3C~58.

If we assume constant velocity expansion, then the age of \GX\ is
$870^{+200}_{-150}$~yr.  This is considerably younger than the
characteristic age of the pulsar, which is 4800~yr.  In fact, the
nominal age of 870~yr is $\sim$100~yr younger then the Crab pulsar and
would make \PSRJ\ the youngest known Galactic pulsar with the possible
exception of PSR J1846$-$0258 in the remnant Kes~75\footnote{No
expansion-speed measurement is available for Kes~75.  The spindown age
of PSR J1846$-$0258 is 723~yr \citep[e.g.,][]{Livingstone+2006}.
Arguments have been advanced that Kes~75, and thus its pulsar, are
considerably older.  These arguments, however, were based in part on a
large distance to Kes~75.  Very recently, \citet{LeahyT2007}, have
shown that a considerably smaller distance is likely for PSR
J1846$-$0258, and thus an age nearer the spindown age is quite
possible.  The other possible exception is PSR J0205+6449 in 3C~58,
which has often been associated with the supernova of 1181 A.D., which
would give it an age of $\sim$825~yr.  A number of arguments, however,
suggest that 3C~58 and PSR J0205+6449 are considerably older
\citep[see for example the discussions in Bietenholz 2006
or][]{Chevalier2005}}\nocite{3C58-2006}.
We note, however, that the expansion of a young PWN is expected
to accelerate slightly as it expands into the freely expanding ejecta for
the first $\sim$1000~yr of its life \citep{vdSwaluw+2001,
BlondinCF2001, Jun1998}.  Such acceleration is in fact observed in the
case of the Crab Nebula \citep{Trimble1968, Nugent1998, Crab-expand}.
If \GX\ is similarly accelerated, then its age would be somewhat older
than 870~yr.  In fact, it would have very nearly the same age as the
Crab nebula, since both the constant-velocity expansion age of both
the Crab's synchrotron nebula \citep{Crab-expand} is almost the same
as that we have determined for \GX.
The only other pulsar known to be younger than this is that in
$0540-69$, whose age is also determined by measuring the expansion
speed of the nebula, albeit from the optical spectrum rather than from
radio imaging \citep{Kirshner+1989}.  With the exception of the pulsar
spindown age, which often differs from the true age of the pulsar by
factors of a few, this is the first age measurement for \GX.

The supernova responsible for creating PSR J1833$-$1034 and \GX\ would
then have exploded around the year 1140~AD.  It might be considered
surprising that no historical supernova was recorded at that time near
\GX, although we note that \citet{Wang+1986} suggested that \GX\ might
be the remnant of the ``guest star'' (supernova) of 48~BC, making it
$\sim$2100~yr old, which is compatible with our measurement only if
strong acceleration has occurred since the supernova explosion.
However, as pointed out by \citet{Camilo+2006}, the hydrogen column
density implied by various X-ray measurements \citep[$\sim 2 \times
10^{22}$~cm$^{-2}$,][]{Warwick+2001, Safi-Harb+2001} in the direction
of \GX\ corresponds to sufficient visual extinction (10--11~mag) for
it to be unlikely that the supernova event was noticed.

We also produced a radio spectral index image, which shows that \GX's
spectrum is fairly uniform, with an average spectral index between 1.5
and 5~GHz of $\alpha = +0.08^{+0.06}_{-0.09}$.  \GX\ is similar in
this regard to two other young PWNe, the Crab and 3C~58, whose radio
spectra have been found to be remarkably uniform over the nebulae.
The spectral uniformity suggests there is a single source for the
relativistic electrons responsible for the radio emission, namely the
pulsar.

One might also expect some radio emission from electrons accelerated
in the shocks produced by the interaction of the expanding supernova
ejecta with their surroundings, and those bounding the pulsar nebula
as it expands into the ejecta.  Such shocks are expected to give rise
to radio spectra steeper than those observed in PWNe (typically with
$\alpha \la -0.5$).  No such regions are seen in \GX.  In particular,
on our image, a region of width $\sim$10\arcsec\ around the edge of
the nebula has a mean spectral index which differed by $<0.03$ from
the mean over the entire nebula.

As mentioned earlier, a faint ``halo'' of $\sim$2.5\arcmin\ in
diameter is seen surrounding the PWN in the X-ray, with about
$\sim$1\% of the latter's surface brightness \citep{Bocchino+2005,
MathesonS2005, Safi-Harb+2001, Slane+2000}.  This halo might be X-ray
emission associated with the outer shock, although its nature is not
yet clear.  Both \citet{MathesonS2005} and \citet{Bocchino+2005} argue
that most of the halo X-ray emission is not from the outer shock, but
is due to dust scattering.

Is there any radio emission associated with this halo? Previous radio
observations have not shown the presence of a corresponding radio halo
to a $1\sigma$ surface brightness limit of 4~\Wmsr{-21}
($\sim$30~mJy~arcmin$^{-2}$) at 1~GHz \citep{BockWD2001, Slane+2000}.
We produced a wide-field radio image from our new radio data to search
for such extended emission, but found no features above the noise
outside the region shown in Fig.~\ref{fg21image}.  We note, however,
that our image background rms of only about 1\% of the brightness of
the PWN is probably not low enough to detect a halo comparable to the
X-ray one, and is in fact somewhat higher than existing limits on the
halo radio emission.
Deeper, low-frequency radio observations of \GX\ should be undertaken
to determine, or set lower limits to, the radio brightness of the
halo.

It has already been pointed out, by \citet{Woltjer+1997} and
\citet{Salvati+1998}, that interpreting the relatively low break
frequency ($\sim$100~GHz) measured in the broadband spectrum of \GX\
\citep[see][]{BockWD2001, GallantT1998, Salter+1989} as being caused
by synchrotron ageing implies either an unrealistically high magnetic
field or an unrealistically large age. For example, if we take the
minimum energy field calculated from synchrotron theory of
$\sim$440~$\mu$G \citep{Slane+2000}, and take the break frequency to be
100~GHz, one would arrive at an age of 13,000~yr
\citep{BockWD2001}.  Our measurement of the expansion speed now
provides firm observational evidence against such large ages.  A much
larger magnetic field, on the other hand, is not feasible on energetic
grounds.  Possibly the break in the spectrum is not due to synchrotron
ageing.  \citet{FleishmanB2007} showed that, in the case of a tangled
magnetic field, several breaks occur in the synchrotron spectrum even
for a single power-law distribution of relativistic electrons, and
these breaks can naturally explain the broadband spectra of PWNe, both
ones like the Crab with break frequencies in the infrared, and ones
like \GX\ or 3C~58, which have much lower break frequencies.  Such a
model might be applicable to \GX\ as well.

Can we identify any radio features associated with the termination
shock in the pulsar wind? The termination shock's radius can be
estimated by balancing the ram pressure of the pulsar wind against the
nebular pressure, and one finds that the radius of the termination
shock is $\sim$1\farcs 6 $\eta$ where $\eta$ is the fractional solid
angle covered by the wind
\citep[see also Slane et al.\ 2000, who obtained a termination
shock angular radius of 1\farcs 5 $\eta$ for input
parameters somewhat different than ours.]{Camilo+2006}
\nocite{Slane+2000}.

The pulsar in \GX\ is surrounded by a region of somewhat lower radio
surface brightness, and the possible elliptical structure mentioned in
\S~\ref{simg} (see Figure~\ref{fcenter}), which has a radius of
$\sim$4\arcsec.  In high-resolution X-ray images, there is a central,
bright region which has an elliptical shape, elongated along the
northwest-southeast direction, with a size of $7\arcsec \times 5
\arcsec$ \citep[see also Matheson \& Safi-Harb 2005; Slane et al.\
2000]{Camilo+2006}\nocite{MathesonS2005, Slane+2000}.  If both these
radio and X-ray structures are interpreted as being associated with
the pulsar wind termination shock, then a radius for the latter of
$\sim$4\arcsec\ is suggested, which implies that $\eta \sim 0.4$.
Given that the only two well-resolved termination shock structures,
being those in the Crab Nebula and in Vela, seem to lie predominately
in the pulsar's equatorial plane, and thus have $\eta < 1$, it seems
this interpretation is not unreasonable.  A more conclusive
interpretation will have to await higher-resolution radio and/or X-ray
observations.

\bibliographystyle{apj}
\bibliography{mybib1,g21.5}

\begin{thebibliography}{43}
\expandafter\ifx\csname natexlab\endcsname\relax\def\natexlab#1{#1}\fi

\bibitem[{{Altenhoff} {et~al.}(1970){Altenhoff}, {Downes}, {Goad}, {Maxwell},
  \& {Rinehart}}]{Altenhoff+1970}
{Altenhoff}, W.~J., {Downes}, D., {Goad}, L., {Maxwell}, A., \& {Rinehart}, R.
  1970, \aaps, 1, 319

\bibitem[{{Becker} \& {Kundu}(1976)}]{BeckerK1976}
{Becker}, R.~H., \& {Kundu}, M.~R. 1976, \apj, 204, 427

\bibitem[{{Becker} \& {Szymkowiak}(1981)}]{BeckerS1981}
{Becker}, R.~H., \& {Szymkowiak}, A.~E. 1981, \apjl, 248, L23

\bibitem[{{Bietenholz}(2006)}]{3C58-2006}
{Bietenholz}, M.~F. 2006, \apj, 645, 1180

\bibitem[{{Bietenholz} {et~al.}(2001{\natexlab{a}}){Bietenholz}, {Frail}, \&
  {Hester}}]{Crab-2001}
{Bietenholz}, M.~F., {Frail}, D.~A., \& {Hester}, J.~J. 2001{\natexlab{a}},
  \apj, 560, 254

\bibitem[{{Bietenholz} {et~al.}(2004){Bietenholz}, {Hester}, {Frail}, \&
  {Bartel}}]{Crab-2004}
{Bietenholz}, M.~F., {Hester}, J.~J., {Frail}, D.~A., \& {Bartel}, N. 2004,
  \apj, 615, 794

\bibitem[{{Bietenholz} {et~al.}(1997){Bietenholz}, {Kassim}, {Frail}, {Perley},
  {Erickson}, \& {Hajian}}]{Crab-1997}
{Bietenholz}, M.~F., {Kassim}, N., {Frail}, D.~A., {Perley}, R.~A., {Erickson},
  W.~C., \& {Hajian}, A.~R. 1997, \apj, 490, 291

\bibitem[{{Bietenholz} {et~al.}(2001{\natexlab{b}}){Bietenholz}, {Kassim}, \&
  {Weiler}}]{3C58-2001}
{Bietenholz}, M.~F., {Kassim}, N.~E., \& {Weiler}, K.~W. 2001{\natexlab{b}},
  \apj, 560, 772

\bibitem[{{Bietenholz} {et~al.}(1991){Bietenholz}, {Kronberg}, {Hogg}, \&
  {Wilson}}]{Crab-expand}
{Bietenholz}, M.~F., {Kronberg}, P.~P., {Hogg}, D.~E., \& {Wilson}, A.~S. 1991,
  \apjl, 373, L59

\bibitem[{{Blondin} {et~al.}(2001){Blondin}, {Chevalier}, \&
  {Frierson}}]{BlondinCF2001}
{Blondin}, J.~M., {Chevalier}, R.~A., \& {Frierson}, D.~M. 2001, \apj, 563, 806

\bibitem[{{Bocchino} {et~al.}(2005){Bocchino}, {van der Swaluw}, {Chevalier},
  \& {Bandiera}}]{Bocchino+2005}
{Bocchino}, F., {van der Swaluw}, E., {Chevalier}, R., \& {Bandiera}, R. 2005,
  \aap, 442, 539

\bibitem[{{Bock} {et~al.}(2001){Bock}, {Wright}, \& {Dickel}}]{BockWD2001}
{Bock}, D.~C.-J., {Wright}, M.~C.~H., \& {Dickel}, J.~R. 2001, \apjl, 561, L203

\bibitem[{{Briggs} {et~al.}(1999){Briggs}, {Schwab}, \&
  {Sramek}}]{BriggsSS1999}
{Briggs}, D.~S., {Schwab}, F.~R., \& {Sramek}, R.~A. 1999, in Astronomical
  Society of the Pacific Conference Series, Vol. 180, Synthesis Imaging in
  Radio Astronomy II, ed. G.~B. {Taylor}, C.~L. {Carilli}, \& R.~A. {Perley},
  127

\bibitem[{{Camilo} {et~al.}(2006){Camilo}, {Ransom}, {Gaensler}, {Slane},
  {Lorimer}, {Reynolds}, {Manchester}, \& {Murray}}]{Camilo+2006}
{Camilo}, F., {Ransom}, S.~M., {Gaensler}, B.~M., {Slane}, P.~O., {Lorimer},
  D.~R., {Reynolds}, J., {Manchester}, R.~N., \& {Murray}, S.~S. 2006, \apj,
  637, 456

\bibitem[{{Chevalier}(2005)}]{Chevalier2005}
{Chevalier}, R.~A. 2005, \apj, 619, 839

\bibitem[{{Davelaar} {et~al.}(1986){Davelaar}, {Smith}, \&
  {Becker}}]{DavelaarSB1986}
{Davelaar}, J., {Smith}, A., \& {Becker}, R.~H. 1986, \apjl, 300, L59

\bibitem[{{Fleishman} \& {Bietenholz}(2007)}]{FleishmanB2007}
{Fleishman}, G.~D., \& {Bietenholz}, M.~F. 2007, \mnras, 376, 625

\bibitem[{{Frail} \& {Moffett}(1993)}]{FrailM1993}
{Frail}, D.~A., \& {Moffett}, D.~A. 1993, \apj, 408, 637

\bibitem[{{F\"urst} {et~al.}(1988){F\"urst}, {Handa}, {Morita}, {Reich},
  {Reich}, \& {Sofue}}]{Furst+1988}
{F\"urst}, E., {Handa}, T., {Morita}, K., {Reich}, P., {Reich}, W., \& {Sofue},
  Y. 1988, \pasj, 40, 347

\bibitem[{{Gaensler} \& {Slane}(2006)}]{GaenslerS2006}
{Gaensler}, B.~M., \& {Slane}, P.~O. 2006, \araa, 44, 17

\bibitem[{{Gallant} \& {Tuffs}(1998)}]{GallantT1998}
{Gallant}, Y.~A., \& {Tuffs}, R.~J. 1998, Memorie della Societa Astronomica
  Italiana, 69, 963

\bibitem[{{Gupta} {et~al.}(2005){Gupta}, {Mitra}, {Green}, \&
  {Acharyya}}]{Gupta+2005}
{Gupta}, Y., {Mitra}, D., {Green}, D.~A., \& {Acharyya}, A. 2005, Current
  Science, 89, 853

\bibitem[{{Jun}(1998)}]{Jun1998}
{Jun}, B.-I. 1998, \apj, 499, 282

\bibitem[{{Kassim}(1992)}]{Kassim1992}
{Kassim}, N.~E. 1992, \aj, 103, 943

\bibitem[{{Kirshner} {et~al.}(1989){Kirshner}, {Morse}, {Winkler}, \&
  {Blair}}]{Kirshner+1989}
{Kirshner}, R.~P., {Morse}, J.~A., {Winkler}, P.~F., \& {Blair}, W.~P. 1989,
  \apj, 342, 260

\bibitem[{{Leahy} \& {Tian}(2007)}]{LeahyT2007}
{Leahy}, D.~A., \& {Tian}, W. 2007, ArXiv e-prints, 711

\bibitem[{{Livingstone} {et~al.}(2006){Livingstone}, {Kaspi}, {Gotthelf}, \&
  {Kuiper}}]{Livingstone+2006}
{Livingstone}, M.~A., {Kaspi}, V.~M., {Gotthelf}, E.~V., \& {Kuiper}, L. 2006,
  \apj, 647, 1286

\bibitem[{{Matheson} \& {Safi-Harb}(2005)}]{MathesonS2005}
{Matheson}, H., \& {Safi-Harb}, S. 2005, Advances in Space Research, 35, 1099

\bibitem[{{Morsi} \& {Reich}(1987)}]{MorsiR1987}
{Morsi}, H.~W., \& {Reich}, W. 1987, \aaps, 69, 533

\bibitem[{{Nugent}(1998)}]{Nugent1998}
{Nugent}, R.~L. 1998, \pasp, 110, 831

\bibitem[{{Safi-Harb} {et~al.}(2001){Safi-Harb}, {Harrus}, {Petre}, {Pavlov},
  {Koptsevich}, \& {Sanwal}}]{Safi-Harb+2001}
{Safi-Harb}, S., {Harrus}, I.~M., {Petre}, R., {Pavlov}, G.~G., {Koptsevich},
  A.~B., \& {Sanwal}, D. 2001, \apj, 561, 308

\bibitem[{{Salter} {et~al.}(1989){Salter}, {Reynolds}, {Hogg}, {Payne}, \&
  {Rhodes}}]{Salter+1989}
{Salter}, C.~J., {Reynolds}, S.~P., {Hogg}, D.~E., {Payne}, J.~M., \& {Rhodes},
  P.~J. 1989, \apj, 338, 171

\bibitem[{{Salvati} {et~al.}(1998){Salvati}, {Bandiera}, {Pacini}, \&
  {Woltjer}}]{Salvati+1998}
{Salvati}, M., {Bandiera}, R., {Pacini}, F., \& {Woltjer}, L. 1998, Memorie
  della Societa Astronomica Italiana, 69, 1023

\bibitem[{{Sault} {et~al.}(1995){Sault}, {Teuben}, \& {Wright}}]{SaultTW1995}
{Sault}, R.~J., {Teuben}, P.~J., \& {Wright}, M.~C.~H. 1995, in Astronomical
  Society of the Pacific Conference Series, Vol.~77, Astronomical Data Analysis
  Software and Systems IV, ed. R.~A. {Shaw}, H.~E. {Payne}, \& J.~J.~E.
  {Hayes}, 433

\bibitem[{{Slane} {et~al.}(2000){Slane}, {Chen}, {Schulz}, {Seward}, {Hughes},
  \& {Gaensler}}]{Slane+2000}
{Slane}, P., {Chen}, Y., {Schulz}, N.~S., {Seward}, F.~D., {Hughes}, J.~P., \&
  {Gaensler}, B.~M. 2000, \apjl, 533, L29

\bibitem[{{Slane} {et~al.}(2004){Slane}, {Helfand}, {van der Swaluw}, \&
  {Murray}}]{Slane+2004}
{Slane}, P., {Helfand}, D.~J., {van der Swaluw}, E., \& {Murray}, S.~S. 2004,
  \apj, 616, 403

\bibitem[{{Tan} \& {Gull}(1985)}]{TanG1985}
{Tan}, S.~M., \& {Gull}, S.~F. 1985, \mnras, 216, 949

\bibitem[{{Trimble}(1968)}]{Trimble1968}
{Trimble}, V. 1968, \aj, 73, 535

\bibitem[{{van der Swaluw} {et~al.}(2001){van der Swaluw}, {Achterberg},
  {Gallant}, \& {T{\'o}th}}]{vdSwaluw+2001}
{van der Swaluw}, E., {Achterberg}, A., {Gallant}, Y.~A., \& {T{\'o}th}, G.
  2001, \aap, 380, 309

\bibitem[{{Wang} {et~al.}(1986){Wang}, {Liu}, {Gorenstein}, \&
  {Zombeck}}]{Wang+1986}
{Wang}, Z.~R., {Liu}, J.~Y., {Gorenstein}, P., \& {Zombeck}, M.~V. 1986,
  Highlights of Astronomy, 7, 583

\bibitem[{{Warwick} {et~al.}(2001){Warwick}, {Bernard}, {Bocchino},
  {Decourchelle}, {Ferrando}, {Griffiths}, {Haberl}, {La Palombara}, {Lumb},
  {Mereghetti}, {Read}, {Schaudel}, {Schurch}, {Tiengo}, \&
  {Willingale}}]{Warwick+2001}
{Warwick}, R.~S., {Bernard}, J.-P., {Bocchino}, F., {Decourchelle}, A.,
  {Ferrando}, P., {Griffiths}, R.~G., {Haberl}, F., {La Palombara}, N., {Lumb},
  D., {Mereghetti}, S., {Read}, A.~M., {Schaudel}, D., {Schurch}, N., {Tiengo},
  A., \& {Willingale}, R. 2001, \aap, 365, L248

\bibitem[{{Wilson} \& {Weiler}(1976)}]{WilsonW1976}
{Wilson}, A.~S., \& {Weiler}, K.~W. 1976, \aap, 49, 357

\bibitem[{{Woltjer} {et~al.}(1997){Woltjer}, {Salvati}, {Pacini}, \&
  {Bandiera}}]{Woltjer+1997}
{Woltjer}, L., {Salvati}, M., {Pacini}, F., \& {Bandiera}, R. 1997, \aap, 325,
  295

\end{thebibliography}

\label{lastpage}

\end{document}